\documentclass[onecolumn,preprintnumbers,amssymb,amsmath,superscriptaddress,letterpaper,nofootinbib]{revtex4}[12pt]

\usepackage{graphicx}
\usepackage{dcolumn}
\usepackage{bm}
\usepackage{natbib}
\usepackage{pstricks}
\usepackage{epsfig}
\usepackage{epstopdf}
\usepackage[toc]{appendix}

\newcommand{\be}{\begin{equation}}
\newcommand{\ee}{\end{equation}}
\newcommand{\bea}{\begin{eqnarray}}
\newcommand{\eea}{\end{eqnarray}}

\begin{document}

\title{Gluing different gravitational  models: $f(R)$ case}

\author{Amin Aalipour}
\email{amin_aalipour@physics.sharif.edu}
\affiliation{Department of Physics, Sharif University of Technology, Tehran 11155-9161, Iran}

\author{Nima Khosravi}
\email{nima@sharif.edu}
\affiliation{Department of Physics, Sharif University of Technology, Tehran 11155-9161, Iran}

\date{\today}

\begin{abstract}
This paper presents a comprehensive analysis of junction conditions for gluing different $f(R)$ gravitational theories across a non-null hypersurface. Using the variational approach, we systematically derive the junction conditions for both general $f(R)$ theories and the special case of Einstein gravity, for comparison. We demonstrate that when joining two distinct $f(R)$ theories, the junction conditions require continuity of  $\partial f(R)/\partial R$, the extrinsic curvature $K_{\mu \nu}$, while allowing for discontinuities in the Ricci Scalar $R$. Furthermore, we establish the equivalence between Jordan and Einstein frame formulations through careful treatment of conformal transformations; Our results reveal that different $f(R)$ theories can be consistently matched provided specific relations between their functional forms and geometric quantities are satisfied at the interface.

\end{abstract}

 \maketitle

\section{Introduction}
The need for modified gravity models has some roots in pursuit of quantum gravity \cite{Basile:2024oms}, the cosmological puzzles e.g. dark energy and dark matter \cite{Sahni:2004ai} and more recently the cosmological anomalies \cite{CosmoVerseNetwork:2025alb}. However, the local observations match the standard Einstein gravity very well which plays a severe role constraining the modified gravity models. A way to bypass this constraint is what is well-known as the screening mechanism \cite{Joyce:2014kja}. In this mechanism, the gravitational force's act depends on the environment e.g. the surrounding density. In these frameworks if the transition be sharp enough one needs the junction condition between two different theories. On the other hand in the \"uber-gravity this transition occurs at a given scalar curvature which is very sharp by construction  \cite{Khosravi:2017aqq,Khosravi:2016kfb}. There are also some models to remove the black-holes singularity by assuming a kind of modified gravity at the center of black holes \cite{Regular}. To study these regular black holes one also needs to know how two different model should be connected \cite{Regular BH,Reg BH,RBH}. Such configurations demand a consistent mathematical framework for joining these disparate space-times across a common boundary, with the governing matching equations known as junction conditions.


The derivation of junction conditions can be approached through three primary methodologies \cite{mansoori}. The \emph{geometric approach} directly analyzes field equations, evaluating potentially singular terms at the boundary—this provides clear geometric insight but becomes cumbersome for theories with higher-order derivatives. The \emph{distributional approach} treats geometric quantities as distributions, with junction conditions emerging from requiring well-defined regular parts of field equations—while rigorous, this method requires careful handling of distribution products. The \emph{variational approach}, employed in this work, applies the principle of least action to systematically yield both bulk field equations and boundary junction conditions simultaneously. Its advantage lies in handling complex field theories in a unified manner, with boundary terms arising naturally from integration by parts.

In this paper, we employ the variational approach to investigate junction conditions for gluing different $f(R)$ theories. Our analysis reveals several key findings: first, we obtain generalized Israel junction conditions for two Einstein gravities with different couplings; second, exhibit novel features for general $f(R)$ theories; third, we carefully address the transformation properties of boundary terms under conformal transformations between Jordan and Einstein frames. These results provide a foundation for constructing composite space-time models where different gravitational theories operate in adjacent regions, with applications to phase transitions in the early universe, interior structures of compact objects, and the interface between classical and quantum gravitational descriptions.
 \section{Gluing $f(R)$ theories}
 Consider two $C^3$ orientable manifolds, $\mathcal{M}^+$ and $\mathcal{M}^-$ each decomposed into two distinct parts along non-null boundaries $\partial \mathcal{M}^+$ and $\partial \mathcal{M}^-$, respectively.  The space-time consists of the union of these distinct sub-manifolds and the boundary of interest, $\mathcal{M}=\mathcal{M^+}\cup \partial{\mathcal{M}} \cup \mathcal{M^-}$, with $x^+,y,x^-$ as established coordinate system respectively\footnote{For the sake of simplicity, $x^\pm$ has been abbreviated to $x$.}. Assuming continuous coordinates 
 on both manifolds at the boundary, yields the so called Lichnerowicz junction condition
  \begin{equation} \label{Lichnerowicz}
 	\begin{split}
 		{g}_{\mu \nu}|_{\partial \mathcal{M}^+}={g}_{\mu \nu}|_{\partial \mathcal{M}^-},
 	\end{split}
 \end{equation}
  which implies the continuity of the induced metric,  ${h}_{\mu \nu}|_{\partial \mathcal{M}^+}  ={h}_{\mu \nu}|_{\partial \mathcal{M}^-}$ that defined as
  \begin{equation} \label{induced metric}
  	{h}_{\mu \nu}= {g}_{\mu \nu} - \epsilon n_\mu n_\nu,
  \end{equation}
 where $n_\mu$ is the unit vector normal to the boundary that is continuous across $\partial{\mathcal{M}}$ and $\epsilon=\pm1$ indicates time-like or space-like surface respectively. The objective of this section is to describe two $f^\pm(R)$ gravity theories with respect to the sub-manifolds $\cal{M}^\pm$ which are separated by a hypersurface $\partial{\cal{M}}$.
 
\subsection{$f(R)=R$ case}
For future comparison and fixing the notation in this sub-section we review the standard junction conditions for the special case of $f(R)=R$ i.e. the Einstein gravity. There is just a modification which is two different Newton constants, $\kappa^\pm$, on both sides of the junction. The total action is consist of the gravitational and the matter part
\begin{equation} \label{SG and SM}
	\begin{split}
		S_G[g]&= \int_\mathcal{M^\pm}  d^4x \sqrt{-g^\pm}\,\frac{1}{2\kappa^\pm}  R^\pm-\frac{1}{ \kappa} \oint_{\partial \mathcal{M}} d^3y \epsilon  \sqrt{|h|} \, K + \int_{\partial \mathcal{V}^\pm} d^3y \epsilon  \sqrt{|h|} \, \frac{K_0^\pm}{ \kappa^\pm} , \\
		S_M[g;\psi]&=\int_\mathcal{M^\pm} d^4x \sqrt{-g^\pm}  L^\pm(g^{\pm}_{\mu \nu};\psi^\pm, \partial_\alpha \psi^\pm)+\oint_{\partial \mathcal{M}} d^3y \sqrt{|h|} \, \mathcal{L}(h_{\mu \nu};\psi , \partial_\alpha \psi),
	\end{split}
    \end{equation}
where $K$ and $\psi$ represent the extrinsic curvature and matter fields respectively. Here the integration on $\mathcal{M^\pm}$ indicates the addition of the integrals over individual sub-manifolds, the last term ($K_0^{\pm}$) indicates a counter-term in order to have a finite action at infinities but it has no effect on the dynamics of the bulk and the boundary\footnote{The non-dynamical term can be extended to a closed integral on the boundary by including the $\partial\cal{M}^\pm$ terms. In that case the junction conditions will be affected by the value of $K_0$. }. If a non-null hypersurface separate the manifold to two sub-manifold, the above boundary integrals decompose correspondingly as
\begin{equation} \label{closed integral}
	\oint_{\partial \mathcal{M}} d^3y \sqrt{|h|} \Psi =	\int_{\partial \mathcal{M}^+} d^3y \sqrt{|h|} \Psi^+ -	\int_{\partial \mathcal{M}^-} d^3y \sqrt{|h|} \Psi^- + 	\int_{\partial \mathcal{V}^\pm} d^3y \sqrt{|h|} \Xi^\pm,
\end{equation}
where $\Psi^\pm$ could be any kind of field on $\cal{M}^\pm$ and the last term indicates the non-gluing boundaries with corresponding fields $\Xi^\pm$. Here the continuity of the metric at the hypersurface has been assumed. It is worthwhile to mention that vanishing \eqref{closed integral} implies vanishing of the each term individually, unless $\partial \mathcal{M}^+=\partial \mathcal{M}^-$. In this case, the first two integral of the \eqref{closed integral} can be combined with $\Psi^+-\Psi^-$ as integrand.

 By the virtue of action principle, the variation of action with respect to $g$ and $\psi$ vanishes. According to \cite{pad}, the variation of the above actions can be written as 
\begin{equation} \label{dSG}
		\delta  S_G[g]= \int_\mathcal{M^\pm} d^4x \sqrt{-g}\, \frac{G_{\mu \nu}^\pm}{2\kappa^\pm} \delta g^{\pm \mu \nu} +\frac{1}{2\kappa} \oint_\mathcal{\partial M} d^3y \epsilon \big(\sqrt{|h|}  \delta I-\delta(2\sqrt{|h|} \, K) \big) - \int_\mathcal{\partial \mathcal{V}^\pm} d^3y \epsilon \sqrt{|h|} \frac{K_0^\pm}{ 2\kappa^\pm} h_{\mu \nu}^\pm \delta h^{\pm \mu \nu} \,  ,
\end{equation}
where 
\begin{equation} \label{DI}
	\delta I\equiv\frac{1}{\sqrt{|h|}} \delta( 2 \sqrt{|h|} K)- (K_{\mu \nu} - K h_{\mu \nu}) \delta h^{\mu \nu}+D_\mu U^\mu,
\end{equation}
with $D_\mu = h_{\mu }^{\ \nu} \nabla_\nu $ and  $U^\mu=2 h^\mu_{ \ \alpha} n_\beta \delta g^{\alpha \beta}-n^\mu  h_{\alpha \beta} \delta h^{\alpha \beta}$. Here the extrinsic curvature defined as \eqref{Extrinsic curvature}. In a similar manner the variation of the matter field has the following form
\begin{equation} \label{dSM}
	\begin{split}
\delta S_M[g;\psi] = &\int_\mathcal{M^\pm} d^4x \sqrt{-g} \{-\frac{1}{2}\tau^\pm_{\mu \nu} \delta g^{\pm \mu \nu} +\big( \frac{\partial {L}^\pm}{\partial \psi^\pm} - \nabla_\alpha \frac{ \partial {L}^\pm}{\partial \partial_\alpha \psi^\pm} \big) \delta \psi^\pm \} 	 \\
+&\oint_\mathcal{\partial M} d^3y \sqrt{|h|} \big\{  -\frac{1}{2}  \Sigma_{\mu \nu}  \delta h^{\mu \nu} +  ( \epsilon n^\alpha \frac{\partial {L}}{ \partial \partial_\alpha \psi} + \frac{\partial \mathcal{L}}{\partial \psi} - D_\alpha \frac{ \partial {\mathcal{L}}}{\partial \partial_\alpha \psi} )\delta \psi     \big\}, \\
-&\oint_{\partial \partial \mathcal{M}} d^2z \sqrt{\sigma} \, \epsilon N_\alpha \frac{ \partial {\mathcal{L}}}{\partial \partial_\alpha \psi} \delta \psi
\end{split}
\end{equation}
where $N^\alpha$ is the unit vector normal to $\partial \partial \mathcal{M} $ and $n_\alpha$. Here
\begin{equation}
	\begin{split}
		\tau_{\mu \nu}^\pm(\psi) &\equiv -2 \frac{\partial {L} }{\partial g^{\mu \nu}} +  {L}  g_{\mu \nu}, \\
		\Sigma_{\mu \nu}(\psi) &\equiv -2 \frac{\partial \mathcal{L} }{\partial h^{\mu \nu}} +  \mathcal{L}  h_{\mu \nu},
	\end{split}
\end{equation}
corresponds the energy momentum tensor of matter field in the bulk and boundary, respectively. Using \eqref{DI}, the total variation can be written as
\begin{equation} \label{dS}
\begin{split}
 \delta S &= \int_\mathcal{M^\pm} d^4x \sqrt{-g} \Big\{\bigg(\frac{G_{\mu\nu}^\pm}{2\kappa^\pm}-\frac{\tau_{\mu \nu}^\pm}{2}\bigg) \delta g^{\pm \mu \nu}+ \bigg( \frac{\partial {L}^\pm}{\partial \psi^\pm} - \nabla_\alpha \frac{ \partial {L}^\pm}{\partial \partial_\alpha \psi^\pm} \bigg) \delta \psi^\pm \Big\} \\
   	&-\oint_\mathcal{\partial M} d^3y \sqrt{|h|} \Big\{\bigg( \frac{\epsilon}{2 \kappa} [K_{\mu \nu}-K h_{\mu \nu}]+\frac{1}{2}  \Sigma_{\mu \nu} \bigg) \delta h^{\mu \nu} - \bigg( \epsilon n_\alpha \frac{\partial {L}}{ \partial \partial_\alpha \psi} + \frac{\partial \mathcal{L}}{\partial \psi} - D_\alpha \frac{ \partial {\mathcal{L}}}{\partial \partial_\alpha \psi} \bigg)\delta \psi     \Big\} \\
  &- \int_\mathcal{\partial \mathcal{V}^\pm} d^3y  \epsilon \sqrt{|h|} \, \frac{K_0^\pm}{2\kappa^\pm}  h_{\mu \nu}^\pm \delta h^{\pm \mu \nu} -\oint_{\partial \partial \mathcal{M}} d^2z \sqrt{\sigma} N_\alpha \bigg(\frac{U^\alpha}{2\kappa} + \epsilon \frac{ \partial {\mathcal{L}}}{\partial \partial_\alpha \psi} \delta \psi \bigg),
\end{split}
\end{equation}
The corner term vanishes if $\partial \partial \mathcal{M}$ is compact or made of piece-wise continuous boundaries like a cylinder; if this assumption is relaxed the total
derivative term is converted into a corner contribution. In addition, the non-dynamical term has no contribution on the bulk and interested boundary equations of motion, thus the corresponding equations for the bulk reduced to
 \begin{equation} \label{Einstein eq}
 	\begin{split}
 		{G_{\mu \nu}^\pm}&={\kappa^\pm} \tau_{\mu \nu}^\pm, \\
 		\frac{\partial {L}^\pm}{\partial \psi^\pm} &= \nabla_\alpha \frac{ \partial {L}^\pm}{\partial \partial_\alpha \psi^\pm}.
 	\end{split}
 \end{equation}
  The boundary equations of motion is a little tricky. Since we are interested in the gluing part of the boundary, we neglect the last term in \eqref{closed integral}.  Due to the continuity of the metric across the hypersurface, and \eqref{delta g},  it can be understood that $\delta h^{\mu \nu }|_{ \partial \mathcal{M}^+}=\delta h^{\mu \nu }|_{ \partial \mathcal{M}^-}$, thus the generalized Israel junction conditions can be demonstrated as\footnote{In addition, one can find the corresponding junction conditions on $\partial \mathcal{V}^\pm$ as well.}
 \begin{equation}\label{JC-GR}
    \frac{K_{\mu \nu}^+ -K^+  h_{\mu \nu}}{ \kappa^+}-\frac{K_{\mu \nu}^- -K^- h_{\mu \nu}}{ \kappa^-}=-\epsilon (\Sigma_{\mu \nu}^+ -\Sigma_{\mu \nu}^-).
 \end{equation}
 Note that in the case of $\kappa^+=\kappa^-$, the above equation will be reduced to the well-known junction condition for the Einstein gravity. 

Depending on the assumption for the continuity of $\psi$ at $\partial \mathcal{M}$, the boundary equation of motion could be different. In general
\begin{equation} \label{matter JC}
	\int_{\partial \mathcal{M}^+} d^3y \sqrt{|h|}  \bigg( \epsilon n_\alpha \frac{\partial {L^+}}{ \partial \partial_\alpha \psi^+} + \frac{\partial \mathcal{L^+}}{\partial \psi^+} - D_\alpha \frac{ \partial {\mathcal{L^+}}}{\partial \partial_\alpha \psi^+}  \bigg) \delta \psi^+    -	\int_{\partial \mathcal{M}^-} d^3y \sqrt{|h|}  \bigg( \epsilon n_\alpha \frac{\partial {L^-}}{ \partial \partial_\alpha \psi^-} + \frac{\partial \mathcal{L^-}}{\partial \psi^-} - D_\alpha \frac{ \partial {\mathcal{L^-}}}{\partial \partial_\alpha \psi^-} \bigg) \delta \psi^-   =0.
\end{equation}
For independent $\psi^\pm$ on the boundary, the terms in the parenthesis independently vanishes \eqref{delta psi}, otherwise $\psi$ is continuous across the boundary thus by $\delta \psi|_{\partial \mathcal{M}^+}=\delta \psi|_{\partial \mathcal{M}^-} $, the integrals can be combined  
\begin{equation} \label{Matter Junction}
 \epsilon n_\alpha \bigg[\frac{\partial {L}^+}{ \partial \partial_\alpha \psi^+}- \frac{\partial {L}^-}{ \partial \partial_\alpha \psi^-}\bigg] + \bigg[\frac{\partial \mathcal{L}^+}{\partial \psi^+}-\frac{\partial \mathcal{L}^-}{\partial \psi^-} \bigg] - \bigg[D_\alpha \frac{ \partial {\mathcal{L}}^+}{\partial \partial_\alpha \psi^+}- D_\alpha \frac{ \partial {\mathcal{L}}^-}{\partial \partial_\alpha \psi^-}\bigg]=0.
\end{equation}
 For the case of smooth matter field i.e. $\mathcal{L}^\pm=0$ and $\Sigma^\pm_{\mu\nu}=0$, the above junction condition reduced to
\begin{equation}\label{JC-GR1}
	\begin{split}
 \dfrac{	K_{\mu \nu}^+ }{\kappa^+} &= \dfrac{	K_{\mu \nu}^- }{\kappa^-} , \\
  n_\alpha \frac{\partial {L}^+}{ \partial \partial_\alpha \psi^+}&=  n_\alpha \frac{\partial {L}^-}{ \partial \partial_\alpha \psi^-},
 	\end{split}
 \end{equation}
 where we have used the trace of (\ref{JC-GR}); which is the result as we had in the Einstein theory of gravity if $\kappa^+ = \kappa^-$.

 \subsection{general $f(R)$ cases}

In this subsection, we study gluing two different $f(R)$ models in mathematical details. We assume there are two sub-manifolds, $\cal{M}^\pm$, separated by a hypersurface $\partial \mathcal{M}$. In each side an $f(R) \in C^2$ model lives, $f^+(R)$ and $f^-(R)$, the gravitational action can be written as \cite{Guarnizo:2010xr}
\begin{equation} \label{fr}
    S_G[g]=\frac{1}{2\kappa} \int_\mathcal{M^\pm} d^4x \sqrt{-g^\pm} \,f^\pm(R^\pm)-\frac{1}{\kappa} \oint_{\partial \mathcal{M}} d^3y \epsilon \sqrt{|h|} \, \frac{\partial f}{\partial R} K + \frac{1}{\kappa} \int_{  \partial \mathcal{V}^\pm} d^3y \epsilon \sqrt{|h|} \, \frac{\partial f^\pm}{\partial R^\pm} K_0^\pm .
\end{equation}
It is well-known that this action can be written as follow with introducing an auxiliary field $\Phi$ 
\begin{equation} \label{SG F}
    S_G[g; \Phi]=\frac{1}{2\kappa} \int_\mathcal{M^\pm} d^4x \sqrt{-g^\pm} (\Phi^\pm R^\pm-U^\pm(\Phi^\pm)) -\frac{1}{\kappa} \oint_{\partial \mathcal{M}} d^3y \epsilon \sqrt{|h|} \, \Phi \, K +\frac{1}{\kappa} \int_{\partial \mathcal{V}^\pm} d^3y \epsilon \sqrt{|h|} \, \Phi^\pm \, K_0^\pm
    \end{equation}
where $U^\pm(\Phi^\pm)\equiv\Phi^\pm R^\pm -f^\pm(\Phi^\pm) $ is the Legendre transformation of $f^\pm(R^\pm)$ such that  $\Phi^\pm \equiv\partial f^\pm(R^\pm)/\partial R^\pm$. This gravity theory, has been coupled to the matter fields $\psi^\pm$, which has be written as \eqref{SG and SM}. The variation of the gravitational terms can been written as follow
\begin{equation} \label{dSG F}
\begin{split}
\delta  S_G[g;\Phi]= &+ \frac{1}{2\kappa} \int_\mathcal{M^\pm} d^4x \sqrt{-g^\pm} \Big\{ \bigg(R^\pm-\frac{\partial U^\pm}{\partial \Phi^\pm}\bigg) \delta \Phi^\pm
+\Big(\Phi^\pm G_{\mu \nu}^\pm+\frac{1}{2} U^\pm g^\pm_{\mu \nu} +\nabla_\alpha \nabla^\alpha \Phi^\pm g^\pm_{\mu \nu}-   \nabla_\mu \nabla_\nu \Phi^\pm\Big)  \delta g^{\pm \mu \nu} \Big\}\\
&+ \frac{1}{2 \kappa} \oint_\mathcal{\partial M} d^3y \epsilon \sqrt{|h|}  \Big\{ \Phi \delta I+\bigg(n_\mu \nabla_\nu \Phi -  n^\alpha \nabla_\alpha \Phi  g_{\mu \nu}\bigg) \delta g^{\mu \nu} -\Phi \frac{1}{\sqrt{|h|}}\delta (2\sqrt{|h|} \, K) -2 K \delta \Phi  \Big\} \\
&+ \frac{1}{2 \kappa} \int_\mathcal{\partial V^\pm} d^3y \epsilon \sqrt{|h|}  \Big\{ 2 K_0^\pm \delta \Phi^\pm  - \Phi^\pm K_0^\pm h_{\mu \nu} \delta h^{\mu \nu}  \Big\},
\end{split}
\end{equation}
where the second term in the boundary integral came by integration by parts. The term $\Phi \delta I$, in \eqref{dSG F} can be expressed as
\begin{equation} \label{FdI}
\begin{split}
   \Phi \delta I =& \frac{ \Phi }{\sqrt{|h|}} \delta(2 \sqrt{|h|} K)-   \Phi (K_{\mu \nu} - K h_{\mu \nu}) \delta h^{\mu \nu}+ \Phi D_\mu U^\mu \\
   =& \frac{ \Phi }{\sqrt{|h|}} \delta(2  \sqrt{|h|} K)-   \Phi (K_{\mu \nu} - K h_{\mu \nu}) \delta h^{\mu \nu}+  D_\mu (\Phi  U^\mu)- U^\mu D_\mu \Phi.
   \end{split}
\end{equation}
Thus the first term can be used to cancel the term in the \eqref{dSG F}, and the second term is a generalization of Gibbons-Hawking-York stress tensor, while the third term can be written as a corner term. The last term can be simplified to
\begin{equation} \label{UDF}
    - U^\mu D_\mu \Phi=- (2 h^\mu_{ \ \alpha} n_\beta \delta g^{\alpha \beta}-n^\mu  h_{\alpha \beta} \delta h^{\alpha \beta}) h_{\mu }^{\ \nu} \nabla_\nu \Phi=-2 h^\alpha_{ \ \mu} n_\nu  \nabla_\alpha \Phi \delta g^{\mu \nu} 
\end{equation}
which can be simplified again with the second term in the boundary integral after some algebras, resulting 
\begin{equation} \label{dS F}
\begin{split}
			\delta  S[g;\Phi,\psi]= &+ \frac{1}{2\kappa} \int_\mathcal{M^\pm} d^4x \sqrt{-g^\pm} \bigg\{ \bigg(R^\pm -\frac{\partial U^\pm}{\partial \Phi^\pm}\bigg) \delta \Phi^\pm
			+ 2 \kappa \bigg(\frac{\partial {L}^\pm}{\partial \psi^\pm} - \partial \frac{ \partial {L}^\pm}{\partial \partial \psi^\pm} \bigg) \delta \psi^\pm \\
			&+\Big(\Phi^\pm G_{\mu \nu}^\pm+\frac{1}{2} U^\pm g^\pm_{\mu \nu} +\nabla_\alpha \nabla^\alpha \Phi^\pm g^\pm_{\mu \nu}-   \nabla_\mu \nabla_\nu \Phi^\pm -\kappa \tau_{\mu \nu }^\pm \Big)  \delta g^{\pm \mu \nu}  \bigg\}  \\
			&- \frac{1}{2 \kappa} \oint_\mathcal{\partial M} d^3y \epsilon \sqrt{|h|} \bigg\{2 K \delta \Phi  + \bigg( \Phi \, [K_{\mu \nu} - K h_{\mu \nu} ] + n^\alpha \nabla_\alpha \Phi h_{\mu \nu}+ \epsilon \kappa  \Sigma_{\mu \nu}\bigg) \delta h^{\mu \nu} \\
			&- h_{\mu}^\alpha  \nabla_\alpha \Phi  \delta n^\mu  -2\kappa \epsilon \bigg(\epsilon n_\alpha \frac{\partial {L}}{ \partial \partial_\alpha \psi} + \frac{\partial \mathcal{L}}{\partial \psi} - D_\alpha \frac{ \partial {\mathcal{L}}}{\partial \partial_\alpha \psi} \bigg)\delta \psi  \bigg\}\\
    &+ \frac{1}{2 \kappa} \int_\mathcal{\partial V^\pm} d^3y \epsilon \sqrt{|h|}  \bigg\{  2 K_0^\pm \delta \Phi^\pm - \Phi^\pm K_0^\pm h_{\mu \nu}^\pm \delta h^{\pm \mu \nu} \bigg\}-\oint_{\partial \partial \mathcal{M}} d^2z \sqrt{\sigma}  N_\alpha \bigg(\frac{\Phi U^\alpha}{2\kappa} + \epsilon \frac{ \partial {\mathcal{L}}}{\partial \partial_\alpha \psi} \delta \psi \bigg).
\end{split}
\end{equation}
It is convenient to express the above variation in terms of variations of $\delta h_{\mu \nu}  $; since due to \eqref{Sigma n}, and $K_{\mu \nu} n^\mu=0$, the $\delta n^\mu$ term can be replaced in favor of $\delta h_{\mu \nu} $ leading 
\begin{equation} \label{variation of f}
	\begin{split}
			\delta  S[g;\Phi,\psi]= &+ \frac{1}{2\kappa} \int_\mathcal{M^\pm} d^4x \sqrt{-g^\pm} \bigg\{ \bigg(R^\pm -\frac{\partial U^\pm}{\partial \Phi^\pm}\bigg) \delta \Phi^\pm
		+ 2 \kappa \bigg(\frac{\partial {L}^\pm}{\partial \psi^\pm} - \partial \frac{ \partial {L}^\pm}{\partial \partial \psi^\pm} \bigg) \delta \psi^\pm  \\
		&+\Big(\Phi^\pm G_{\mu \nu}^\pm+\frac{1}{2} U^\pm g^\pm_{\mu \nu} +\nabla_\alpha \nabla^\alpha \Phi^\pm g^\pm_{\mu \nu}-   \nabla_\mu \nabla_\nu \Phi^\pm -\kappa \tau_{\mu \nu }^\pm \Big)  \delta g^{\pm \mu \nu}  \bigg\}  \\
		&+ \frac{1}{2 \kappa} \oint_\mathcal{\partial M} d^3y \epsilon \sqrt{|h|} \bigg\{ \bigg( \Phi \, [K^{\mu \nu} - K h^{\mu \nu} ] +  n^\alpha \nabla_\alpha \Phi h^{\mu \nu} -  n^\mu h^{\nu \alpha}  \nabla_\alpha \Phi + \epsilon \kappa  \Sigma^{\mu \nu}\bigg) \delta h_{\mu \nu} -2 K \delta \Phi \\
		&+  2\kappa \epsilon (\epsilon n_\alpha \frac{\partial {L}}{ \partial \partial_\alpha \psi} + \frac{\partial \mathcal{L}}{\partial \psi} - D_\alpha \frac{ \partial {\mathcal{L}}}{\partial \partial_\alpha \psi} )\delta \psi  \bigg\} + \frac{1}{2 \kappa} \int_\mathcal{\partial V^\pm} d^3y \epsilon \sqrt{|h|}  \ K_0^\pm \bigg\{  2  \delta \Phi^\pm + \Phi^\pm  h^{\pm \mu \nu} \delta h_{\mu \nu}^\pm \bigg\}\\
		&-\frac{1}{2\kappa} \oint_{\partial \partial \mathcal{M}} d^2z \sqrt{\sigma}  N_\alpha \bigg(\Phi U^\alpha + 2\kappa \epsilon \frac{ \partial {\mathcal{L}}}{\partial \partial_\alpha \psi} \delta \psi \bigg).
	\end{split}
\end{equation}
The corresponding equation of motion for the matter field has been written in the previous section while for the auxiliary field $\Phi$
\begin{equation} \label{R= chi}
	 R^\pm=\frac{\partial U^\pm}{\partial \Phi^\pm},
\end{equation}
which satisfies trivially  by using the definition of $U^\pm(\Phi^\pm)$; The gravity sector equations are as follows
\begin{equation} \label{F EOM}
    \begin{split}
   \Phi^\pm R_{\mu \nu}^\pm
   +\bigg(  \nabla_\alpha \nabla^\alpha \Phi^\pm  -\frac{1}{2} f^\pm   \bigg) g^\pm_{\mu \nu}- 
    \nabla_\mu \nabla_\nu \Phi^\pm  = \kappa \tau_{\mu \nu }^\pm   .
    \end{split}
\end{equation}
The trace of the above equation yields
\begin{equation} \label{KG for PHI}
	3  \nabla_\alpha \nabla^\alpha \Phi^\pm + \Phi^\pm R^\pm -2 f^\pm(R^\pm)= \kappa \tau^\pm,
\end{equation}
which is a scalar field equation for $\Phi^\pm$ indicating that the auxiliary field has its own independent dynamics\footnote{In fact as it is shown in Appendix C, the dynamical equation for the (extra) scalar degree of freedom in the Einstein frame will be equation (\ref{KG for PHI}) after the conformal transformation.}.  The corresponding junction conditions of the tensor part of this theory can be found as
\begin{equation} \label{JC-F}
	\begin{split}
		 \Big[ K^+_{ \mu \nu} - K^+  h_{\mu \nu} \Big] \Phi^+\Big|_{\partial \mathcal{M}^+}
		- &\Big[K^-_{ \mu \nu} - K^-  h_{\mu \nu}\Big] \Phi^-\Big|_{\partial \mathcal{M}^-}
		+ n^\alpha  \Big[ \nabla_\alpha \Phi^+ -  \nabla_\alpha \Phi^- \Big] h_{\mu \nu} \\
		 - n_\mu h_\nu^{ \alpha}  &\Big[  \nabla_\alpha \Phi^+ -  \nabla_\alpha \Phi^-\Big] =- \epsilon \kappa  \Big[\Sigma^+_{ \mu \nu}- \Sigma^-_{ \mu \nu}\Big].
	\end{split}
\end{equation}
Here the equation has been written in covariant form. Taking the product of $n^\mu$ with \eqref{JC-F}, leads to
\begin{equation} \label{jc del R}
	h_\nu^{ \alpha}  \Big[  \nabla_\alpha \Phi^+ -  \nabla_\alpha \Phi^-\Big] =0.
\end{equation}
This shows that the term corresponding to $\delta n^\mu$ is indeed a separate constraint.  The junction condition for the matter field is as previous section. Depending to the continuity of the $\Phi$, similar to the matter field, the auxiliary field may have jump across the hypersurface. Generally the junction condition for $\Phi$,
\begin{equation} \label{JC F chi}
	 \int_\mathcal{\partial M^+} d^3y \epsilon \sqrt{|h|} \,	K^+  \delta \Phi^+  -  \int_\mathcal{\partial M^-} d^3y \epsilon \sqrt{|h|} \, 	K^-   \delta \Phi^-=0,
\end{equation}
implies vanishing of the integrands separately; while considering the continuity of the auxiliary field across the boundary,   $\Phi^+|_{\partial \mathcal{M^+}}= \Phi^-|_{\partial \mathcal{M^-}} \equiv \Phi|_{\partial \mathcal{M}} $ leading to  $\delta \Phi|_{\partial \mathcal{M^+}}=  \delta \Phi|_{\partial \mathcal{M^-}}$ hence
\begin{equation} \label{Junction conditions R}
	K^+ =	K^- .
\end{equation}
In addition, since tangential derivative commutes with the jump of any quantity, equation (\ref{jc del R}) satisfies trivially, meaning that it isn't an independent condition in a smoothly defined theory. By using the results of the above, 	equation ($\ref{JC-F}$) can be simplified to,
 \begin{equation} \label{REDUCED }
 	 \Big[ K^+_{ \mu \nu}   
 	- K^-_{ \mu \nu} \Big] \Phi\Big|_{\partial \mathcal{M}}
 	+ n^\alpha  \Big[ \nabla_\alpha \Phi^+ -  \nabla_\alpha \Phi^- \Big] h_{\mu \nu} 
  =- \epsilon \kappa  \Big[\Sigma^+_{ \mu \nu}- \Sigma^-_{ \mu \nu}\Big].
 \end{equation}
 Equations (\ref{F EOM}), (\ref{KG for PHI}), (\ref{jc del R}) and (\ref{REDUCED }) can be written in terms of $f(R)$ and its derivatives as well; However, due to consistency with the next section, we proceed with the auxiliary field $\Phi$.  
 
 The interesting consequence of the corresponding junction conditions is the existence of discontinuity in the Ricci scalar. In that case the continuity of $\Phi$ does not necessary implies the smoothness of $R$ across the boundary,
 \begin{equation} \label{Ricci}
 	\begin{split}
 \frac{\partial f^+}{\partial R^+}\bigg|_{\partial \mathcal{M}^+} &=\frac{\partial f^-}{\partial R^-}\bigg|_{\partial \mathcal{M}-} , \\  
 \frac{\partial^2 f^+}{\partial R^{+ 2}}\bigg|_{\partial \mathcal{M}^+} \delta R^+  &=\frac{\partial^2 f^-}{\partial R^{-2}}\bigg|_{\partial \mathcal{M}-} \delta R^- .
\end{split}
 \end{equation} 
 Presuming that $R^+= R^-$ across the boundary, the above equation implies additional constraints. 

\subsubsection{Smooth transition}
The junction conditions for a smooth joining of two gravity theories at the hypersurface $\partial \mathcal{M}$, can be found by setting $\Sigma^\pm_{\mu \nu}=0$. The trace of (\ref{REDUCED }) leads 
\begin{equation} \label{trace}
 n^\alpha  \Big[ \nabla_\alpha \Phi^+ -  \nabla_\alpha \Phi^- \Big] =0.
\end{equation}
According to (\ref{Jump of del S}), the continuity of $\Phi$ implies 
\begin{equation} \label{JC F'}
 \nabla_\alpha \Phi^+ =  \nabla_\alpha \Phi^- , 
\end{equation}
 across the boundary, therefore (\ref{REDUCED }) leads to 
\begin{equation} \label{JC K}
	K^+_{\mu \nu}   =	K^-_{\mu \nu} ,
\end{equation}
The consequence of different gravity theories indicates that even for a smooth transition, the difference of Ricci scalar does not vanish in general. It has been inferred that different $\Phi$ implies non-equal scalar curvature across the boundary as demonstrated in (\ref{Ricci}), which is another feature of this theory.  

\subsubsection{Standard $f(R)$ theory}
In the case of standard $f(R)$ theory, due to the differentiability of $f(R)$, the function $\Phi^+=\Phi^- \equiv \Phi $ thus if the function $\Phi$ is invertible, then $R^+=R^-$ at the boundary and the second equation of (\ref{Ricci}) implies $\frac{\partial^2 f}{\partial R^{+2}}= \frac{\partial^2 f}{\partial R^{-2}} \equiv \frac{\partial^2 f}{\partial R^2}$, hence equation (\ref{REDUCED }), reduced to 
\begin{equation}
\begin{split}
 \Big[ K^+_{ \mu \nu}   
- K^-_{ \mu \nu} \Big] \frac{\partial f}{\partial R}
+ n^\alpha  \Big[ \nabla_\alpha R^+ -  \nabla_\alpha R^- \Big] \frac{\partial^2 f}{\partial R^2} \  h_{\mu \nu} 
=- \epsilon \kappa  \Big[\Sigma^+_{ \mu \nu}- \Sigma^-_{ \mu \nu}\Big],
\end{split}
\end{equation}
which is consistent with \cite{mansoori}. 
\section{Einstein frame}
An alternative approach to determine the junction condition of two different $f(R)$ gravities, is to work in the Einstein frame. In this frame, the total action can be decomposed as 
\begin{equation} \label{action decomposition}
	\tilde{S}[\tilde{g};\phi,\psi]=\tilde{S}_G[\tilde{g}]+\tilde{S}_F[\tilde{g};\phi]
	+\tilde{S}_M[\tilde{g};\phi,\psi],
\end{equation}
where $\tilde{S}_G$ is simply Einstein-Hilbert gravitational action followed by Gibbons-Hawking boundary term
\begin{equation} \label{SG}
	\tilde{S}_G[\tilde{g}]=\frac{1}{2\kappa} \int_\mathcal{M^\pm} d^4x \sqrt{-\tilde{g}^\pm}\, \tilde{R}^\pm-\frac{1}{ \kappa} \oint_{\partial \mathcal{M}} d^3y \epsilon \sqrt{|\tilde{h}|} \tilde{K} +\frac{1}{ \kappa} \int_{\partial \mathcal{V^\pm}} d^3y \epsilon \sqrt{|\tilde{h}|} \tilde{K}_0^\pm,
\end{equation}
 $\tilde{S}_M$ represent matter fields
 \begin{equation} \label{SM}
 	\tilde{S}_M[\tilde{g};\phi,\psi]=\int_\mathcal{M^\pm} d^4x \sqrt{-\tilde{g}^\pm}  \tilde{L}^\pm(\tilde{g}_{\mu \nu}^\pm;\phi,\psi^\pm, \partial_\alpha \psi^\pm)+\oint_{\partial \mathcal{M}} d^3y \sqrt{|\tilde{h}|} \tilde{\mathcal{L}}(\tilde{h}_{\mu \nu};\phi,\psi, \partial_\alpha \psi),
 \end{equation}
  while $\tilde{S}_F$ is the scalar field action which connects the Einstein frame to the Jordan one,
 \begin{equation} \label{SF}
	\tilde{S}_F[\tilde{g};\phi]=\int_\mathcal{M^\pm} d^4x \sqrt{-\tilde{g}^\pm} \,  \bigg(-\frac{1}{2} \tilde{\nabla}_\alpha \phi^\pm \tilde{\nabla}^\alpha \phi^\pm -V^\pm(\phi^\pm)\bigg)+ \oint_{\partial \mathcal{M}} d^3y \sqrt{|\tilde{h}|} \tilde{\mathfrak{L}}(\tilde{h}_{\mu \nu};\phi),
\end{equation}
where the boundary integral in both matter and scalar field actions, represents surface fields.
  
 By the virtue of the variational principle and the results of the previous section, corresponding action is as follows
 \begin{equation} \label{variation}
 	\begin{split}
 		\delta \tilde{S} = &\int_\mathcal{M^\pm} d^4x \sqrt{-\tilde{g}^\pm} \bigg\{ \bigg(\frac{\tilde{G}_{\mu \nu}^\pm}{2\kappa}-\frac{\tilde{T}_{\mu \nu}^\pm+\tilde{\tau}_{\mu \nu}^\pm}{2} \bigg) \delta \tilde{g}^{\pm \mu \nu }+ \bigg(\tilde{\nabla}_\alpha \tilde{\nabla}^\alpha \phi^\pm-\frac{\partial V^\pm}{\partial \phi^\pm}  + \frac{\partial {\tilde{L}}^\pm}{\partial \phi^\pm} \bigg) \delta \phi^\pm+ \bigg( \frac{\partial {\tilde{L}}^\pm}{\partial \psi^\pm} -  \tilde{\nabla}_\alpha \frac{ \partial {\tilde{L}}^\pm}{\partial \partial_\alpha \psi^\pm} \bigg) \delta \psi^\pm \bigg\} \\
 		-&\oint_\mathcal{\partial M} d^3y \sqrt{|\tilde{h}|} \bigg\{\bigg( \frac{\epsilon}{2 \kappa} [\tilde{K}_{\mu \nu}-\tilde{K} \tilde{h}_{\mu \nu}]+\frac{1}{2} [\tilde{S}_{\mu \nu}+ \tilde{\Sigma}_{\mu \nu} ]\bigg) \delta \tilde{h}^{\mu \nu}-    \bigg(\frac{\partial  \tilde{\mathfrak{L}}}{\partial \phi}  +\frac{\partial {\tilde{\mathcal{L}}}}{\partial \phi}-\epsilon \tilde{n}^\alpha \tilde{\nabla}_\alpha \phi \bigg) \delta \phi \\
 		-& \bigg( \frac{\partial \tilde{\mathcal{L}}}{\partial \psi} - \tilde{D}_\alpha \frac{ \partial {\tilde{\mathcal{L}}}}{\partial \partial_\alpha \psi}+ \epsilon \tilde{n}_\alpha \frac{\partial {\tilde{L}}}{ \partial \partial_\alpha \psi} \bigg)\delta \psi     \bigg\} -\frac{1}{2 \kappa} \int_\mathcal{\partial V^\pm} d^3y \epsilon \sqrt{|\tilde{h}|} \,  {\tilde{K}_0^\pm}  \tilde{h}_{\mu \nu}^\pm \delta \tilde{h}^{\pm \mu \nu}\\
 		-&\oint_{\partial \partial \mathcal{M}} d^2z \sqrt{\tilde{\sigma}} \tilde{N}_\alpha \bigg(\frac{\tilde{U}^\alpha}{2\kappa} + \epsilon \frac{ \partial {\mathcal{\tilde{L}}}}{\partial \partial_\alpha \psi} \delta \psi \bigg)
 	\end{split}
 \end{equation}
 where
 \begin{equation} \label{EMT Bulk}
 	\begin{split}
 	 \tilde{T}_{\mu \nu}^\pm(\phi)  &= \nabla_\mu  \phi^\pm \nabla_\nu \phi^\pm -\tilde{g}_{\mu \nu}^\pm  
 	\bigg( \frac{1}{2}  \tilde{\nabla}_\alpha  \phi^\pm \tilde{\nabla}^\alpha \phi^\pm + V(\phi^\pm) \bigg) ,\\
 	 \tilde{S}_{\mu \nu}(\phi) &=-2 \frac{\partial \tilde{\mathfrak{L}} }{\partial \tilde{h}^{\mu \nu}} +  \tilde{\mathfrak{L}}  \tilde{h}_{\mu \nu},
 	\end{split}
 \end{equation}
 is the energy momentum tensor of the corresponding bulk and boundary scalar field and
 \begin{equation} \label{EMT Boundary}
 	\begin{split}
 	\tilde{\tau}_{\mu \nu}^\pm(\psi) &= -2 \frac{\partial {\tilde{L}} }{\partial \tilde{g}^{\mu \nu}} +  {\tilde{L}}  \tilde{g}_{\mu \nu}, \\
 	 \tilde{\Sigma}_{\mu \nu}(\psi) &= -2 \frac{\partial \tilde{\mathcal{L}} }{\partial \tilde{h}^{\mu \nu}} +  \tilde{\mathcal{L}}  \tilde{h}_{\mu \nu},
 	\end{split}
 \end{equation}
 corresponds the energy momentum tensor of the bulk matter and boundary respectively. The corner term vanishes if $\partial \partial \mathcal{M}$ is compact or made of piece-wise continuous boundaries like a cylinder, hence the equations of motion of the bulk is simply the Einstein's equation, Klein-Gordon equation with a potential, and Euler-Lagrange equations for matter field $\psi$
 \begin{equation} \label{EOM bulk Einstein frame}
 	\begin{split}
 		{\tilde{G}_{\mu \nu}^\pm}&=\kappa \big({\tilde{T}_{\mu \nu}^\pm+\tilde{\tau}_{\mu \nu}^\pm}\big) \, ,\\
 		\tilde{\nabla}_\alpha \tilde{\nabla}^\alpha \phi^\pm&=\frac{\partial V^\pm}{\partial \phi^\pm}-\frac{\partial {\tilde{L}}^\pm}{\partial \phi^\pm} \, ,\\
 		 \frac{\partial {\tilde{L}}^\pm}{\partial \psi^\pm} &=\tilde{\nabla}_\alpha \frac{ \partial {\tilde{L}}^\pm}{\partial \partial_\alpha \psi^\pm} \,.
 	\end{split}
 \end{equation}
 Due to the continuity of the metric across the hypersurface, $\delta \tilde{h}^{\mu \nu }|_{ \partial \mathcal{M}^+}=\delta \tilde{h}^{\mu \nu }|_{ \partial \mathcal{M}^-}$,  the metric junction condition becomes
  \begin{equation} \label{JC Einsten K}
  	[\tilde{K}_{\mu \nu}^+ -\tilde{ K}_{\mu \nu}^-] -[\tilde{K}^+ -\tilde{K}^- ] \tilde{h}_{\mu \nu}= -\epsilon \kappa (\tilde{S}_{\mu \nu}^+ -\tilde{S}_{\mu \nu}^-+ \tilde{\Sigma}_{\mu \nu}^+ - \tilde{\Sigma}_{\mu \nu}^-).
  \end{equation}
   Depending the continuity of $\psi$, the matter field junction condition is the same as (\ref{matter JC}) or (\ref{Matter Junction}); except that one must use the Einstein frame Lagrangians instead of the Jordan frame ones. Concerning the boundary equation of motion for the scalar field, requires the development of continuity for $\phi$, which has been argued in (\ref{Omega}); resulting $\delta \phi|_{ \partial \mathcal{M}^+}=\delta \phi|_{ \partial \mathcal{M}^-}$, thus
   \begin{equation} \label{JC phi }
      \frac{\partial  \tilde{\mathfrak{L}}}{\partial \phi} +\frac{\partial {\tilde{\mathcal{L}}}}{\partial \phi} -\epsilon \tilde{n}^\alpha \tilde{\nabla}_\alpha \phi=0.
   \end{equation}
 These bulk and boundary equations, clearly reduced to the standard equations when $\phi=0$.
  
 A smooth transition across the $\partial \mathcal{M}$ requires vanishing the boundary terms in matter and scalar field action, therefore 
 \begin{equation} \label{JC smooth}
 	\begin{split}
 		\tilde{K}_{\mu \nu}^+ &=	\tilde{K}_{\mu \nu}^- , \\
 		\tilde{\nabla}_\alpha \phi^+ &= \tilde{\nabla}_\alpha \phi^-  . \\
 	\end{split}
 \end{equation}
  where (\ref{Jump of del S}) has been used for the second line.
 
\subsection{Invariance of variations}
One can transform the metric, as $\tilde{g}_{\mu\nu} = \Omega^2\, g_{\mu\nu}$, in the variation of action, and find out that the boundary term reduced to 
\begin{equation} \label{dS tilde}
	\begin{split}
	\delta	\tilde{S} &= \frac{1}{2 \kappa} \int_\mathcal{M^\pm} d^4x \sqrt{-g^\pm}\,   \bigg\{ 	\bigg( 2 \Omega^{\pm } R^\pm -12 \nabla_\alpha \nabla^\alpha\Omega^\pm + 2 \kappa \bigg[   (\Omega^{\pm })^3 \tilde{\tau} -\Omega^\pm \nabla_\alpha \phi^\pm \nabla^\alpha \phi^\pm -4  V^\pm (\Omega^{\pm })^3 \bigg] \bigg) \delta \Omega^\pm \\
	  &+2 \kappa \bigg(  (\Omega^\pm)^{2} \nabla_\alpha \nabla^\alpha \phi^\pm+ 2\Omega^\pm \nabla_\alpha \phi^\pm \nabla^\alpha \Omega^\pm+ (\Omega^{\pm })^4 \bigg[\frac{\partial {\tilde{L}}^\pm}{\partial \phi^\pm}- \frac{\partial V^\pm }{\partial \phi^\pm} \bigg] \bigg) \delta \phi^\pm + 2 \kappa (\Omega^{\pm })^4 \bigg( \frac{\partial {\tilde{L}}^\pm}{\partial \psi^\pm} - \tilde{\nabla}_\alpha \frac{ \partial {\tilde{L}}^\pm}{\partial \partial_\alpha \psi^\pm} \bigg) \delta \psi^\pm    \\
	&+ \bigg(   (\Omega^\pm)^2 \big[ G_{\mu \nu} -\kappa \tilde{\tau}_{\mu \nu} \big] -2 (\Omega^\pm) \nabla_\mu \nabla_\nu \Omega^\pm -2  \nabla_\mu \Omega^\pm  \nabla_\nu \Omega^\pm + \bigg[2 \Omega^\pm   \nabla_\alpha \nabla^\alpha \Omega^\pm +2  \nabla_\alpha \Omega^\pm  \nabla^\alpha \Omega^\pm+
	 \kappa (\Omega^\pm)^4 V^\pm \bigg] g_{\mu \nu}  \bigg) \delta g^{\pm \mu \nu} \bigg\} \\
	&-\frac{1}{2 \kappa}  \oint_\mathcal{\partial M} d^3y \epsilon \sqrt{|h|}  \bigg\{ \bigg(   \Omega^2 \big[K_{\mu \nu}-K h_{\mu \nu} \big] +2  \Omega n^\alpha \nabla_\alpha \Omega  \ h_{\mu \nu} +\kappa   \epsilon \Omega \big[\tilde{S}_{\mu \nu}+ \tilde{\Sigma}_{\mu \nu} \big]\bigg) \delta h^{\mu \nu}+2 \kappa \Omega^2 \bigg( {n}^\alpha {\nabla}_\alpha \phi  - \epsilon \Omega  \bigg[\frac{\partial  \tilde{\mathfrak{L}}}{\partial \phi}  +\frac{\partial {\tilde{\mathcal{L}}}}{\partial \phi}\bigg] \bigg) \delta \phi  \\
	&+    \bigg(    4 \Omega K- 12 n^\alpha \nabla_\alpha \Omega -  2 \kappa \epsilon \Omega^3 \bigg [ \frac{\partial  \tilde{\mathfrak{L}}}{\partial \Omega} +\frac{\partial  \tilde{\mathcal{L}}}{\partial \Omega}\bigg]- 2 \kappa \epsilon \Omega^2 \big[\tilde{S}+ \tilde{\Sigma} \big] \bigg) \delta \Omega       -2 \kappa \Omega^3 \bigg(\epsilon \bigg[ \frac{\partial \tilde{\mathcal{L}}}{\partial \psi} - \tilde{D}_\alpha \frac{ \partial {\tilde{\mathcal{L}}}}{\partial \partial_\alpha \psi} \bigg]+   {n}_\alpha \frac{\partial {\tilde{L}}}{ \partial \partial_\alpha \psi} \bigg)\delta \psi     \bigg\} \\
			&-\frac{1}{2 \kappa} \int_\mathcal{\partial V^\pm} d^3y \epsilon \sqrt{|{h}|} \,   {{K}_0^\pm}  \bigg[(\Omega^\pm)^2 {h}_{\mu \nu}^\pm  \delta {h}^{\pm \mu \nu}-6 \Omega^\pm \delta \Omega^\pm\bigg]
			-\frac{1}{2 \kappa} \oint_{\partial \partial \mathcal{M}} d^2z \sqrt{{\sigma}}  \, \Omega^3  {N}_\alpha \bigg(\Omega^{-1} {U}^\alpha + 2\kappa \epsilon \frac{ \partial {\mathcal{\tilde{L}}}}{\partial \partial_\alpha \psi} \delta \psi \bigg)
	\end{split}
\end{equation}
where we just have scaled $\tilde{K}_0^\pm= (\Omega^\pm)^{-1} K_0^\pm$  and used $N_\alpha n^\alpha=0$ in the corner integral. Comparing the above result with (\ref{variation of f}), it is clear that now we have $g_{\mu\nu}$, $\psi$, $\phi$ and $\Omega$ as our degrees of freedom instead of $g_{\mu\nu}$, $\psi$ and $\Phi$. So to equalize the two approaches the easy guess is the dependency of $\Phi$ to $\phi$ and $\Omega$. This also means $\phi$ and $\Omega$ are not independent. Therefore in the first step let $\phi^\pm = A^\pm(\Omega^\pm)$, by comparing it can be understood that the terms including the covariant derivatives in the bulk action must vanish
\begin{equation} \label{A(omega)}
\bigg[	2 \kappa\, (\Omega^\pm)^{2}\, (\frac{\partial A^\pm}{\partial \Omega^\pm})^2 -12 \bigg] \nabla_\alpha \nabla^\alpha\Omega^\pm + 2 \kappa \, \Omega^\pm\, \frac{\partial A^\pm}{\partial \Omega^\pm} \bigg[  \Omega^\pm  \, \frac{\partial^2 A^\pm}{\partial \Omega^{\pm 2}}+ \frac{\partial A^\pm}{\partial \Omega^\pm} \bigg]  \nabla_\alpha \Omega^\pm \nabla^\alpha \Omega^\pm =0.
\end{equation}
The terms in the bracket must vanish independently leading to $A^\pm(\Omega^\pm)=\sqrt{6/\kappa} \ln \Omega^\pm$ i.e.
\begin{equation} \label{phi A}
\phi^\pm = \sqrt{\frac{6}{\kappa}} \ln \Omega^\pm.	
\end{equation} 
With this choice, (\ref{dS tilde}) reduced to
\begin{equation} \label{red dS tilde}
	\begin{split}
		\delta	\tilde{S} &= \frac{1}{2 \kappa} \int_\mathcal{M^\pm} d^4x \sqrt{-g^\pm}\,   \bigg\{ 	\bigg( 2 \Omega^{\pm } R^\pm + 2 \kappa \bigg[   (\Omega^{\pm })^3 \tilde{\tau} + (\Omega^{\pm })^4 \frac{\partial {\tilde{L}}^\pm}{\partial \Omega^\pm}- \frac{\partial \big( (\Omega^\pm)^4 V^\pm \big) }{\partial \Omega^\pm} \bigg] \bigg) \delta \Omega^\pm + 2 \kappa (\Omega^{\pm })^4 \bigg( \frac{\partial {\tilde{L}}^\pm}{\partial \psi^\pm} - \tilde{\nabla}_\alpha \frac{ \partial {\tilde{L}}^\pm}{\partial \partial_\alpha \psi^\pm} \bigg) \delta \psi^\pm    \\
		&+ \bigg(   (\Omega^\pm)^2 \big[ G_{\mu \nu} -\kappa \tilde{\tau}_{\mu \nu} \big] -2 (\Omega^\pm) \nabla_\mu \nabla_\nu \Omega^\pm -2  \nabla_\mu \Omega^\pm  \nabla_\nu \Omega^\pm + \bigg[2 \Omega^\pm   \nabla_\alpha \nabla^\alpha \Omega^\pm +2  \nabla_\alpha \Omega^\pm  \nabla^\alpha \Omega^\pm+
		\kappa (\Omega^\pm)^4 V^\pm \bigg] g_{\mu \nu}  \bigg) \delta g^{\pm \mu \nu} \bigg\} \\
		&-\frac{1}{2 \kappa}  \oint_\mathcal{\partial M} d^3y \epsilon \sqrt{|h|}  \bigg\{ \bigg(   \Omega^2 \big[K_{\mu \nu}-K h_{\mu \nu} \big] +2  \Omega n^\alpha \nabla_\alpha \Omega  \ h_{\mu \nu} +\kappa   \epsilon \Omega \big[\tilde{S}_{\mu \nu}+ \tilde{\Sigma}_{\mu \nu} \big]\bigg) \delta h^{\mu \nu} \\
		&+    \bigg(    4 \Omega K -  4 \kappa \epsilon \Omega^3 \bigg [ \frac{\partial  \tilde{\mathfrak{L}}}{\partial \Omega} +\frac{\partial  \tilde{\mathcal{L}}}{\partial \Omega}\bigg]- 2 \kappa \epsilon \Omega^2 \big[\tilde{S}+ \tilde{\Sigma} \big] \bigg) \delta \Omega       -2 \kappa \Omega^3 \bigg(\epsilon \bigg[ \frac{\partial \tilde{\mathcal{L}}}{\partial \psi} - \tilde{D}_\alpha \frac{ \partial {\tilde{\mathcal{L}}}}{\partial \partial_\alpha \psi} \bigg]+   {n}_\alpha \frac{\partial {\tilde{L}}}{ \partial \partial_\alpha \psi} \bigg)\delta \psi     \bigg\} \\
		&-\frac{1}{2 \kappa} \int_\mathcal{\partial V^\pm} d^3y \epsilon \sqrt{|{h}|} \,   {{K}_0^\pm}  \bigg[(\Omega^\pm)^2 {h}_{\mu \nu}^\pm  \delta {h}^{\pm \mu \nu}-6 \Omega^\pm \delta \Omega^\pm\bigg]
		-\frac{1}{2 \kappa} \oint_{\partial \partial \mathcal{M}} d^2z \sqrt{{\sigma}}  \, \Omega^3  {N}_\alpha \bigg(\Omega^{-1} {U}^\alpha + 2\kappa \epsilon \frac{ \partial {\mathcal{\tilde{L}}}}{\partial \partial_\alpha \psi} \delta \psi \bigg) .
	\end{split}
\end{equation}
Now both the frames have the same number of degrees of freedom so we need to look for the relation between $\Omega$ and $\Phi$. In addition, the Lagrangians in the conformal frame must be transform to the Jordan frame ones after transforming the scalar field.  In order to achieve this transformation, one must impose $(\Omega^{\pm })^3 \tilde{\tau}+  (\Omega^\pm)^4 \frac{\partial {\tilde{L}}^\pm}{\partial \Omega^\pm}=0$, which by using the definition of $\tilde{\tau}$ reduced to
\begin{equation} \label{imposed eq}
(\Omega^\pm)^4 \frac{\partial {\tilde{L}}^\pm}{\partial \Omega^\pm} +4 (\Omega^{\pm })^3 {\tilde{L}} 	-2(\Omega^{\pm })^3  \tilde{g}^{\mu \nu}  \frac{\partial {\tilde{L}} }{\partial \tilde{g}^{\mu \nu}}  =0.
\end{equation}
The solution of this quasi-linear PDE, illustrates the change of the matter Lagrangian under conformal transformation
\begin{equation} \label{L TILDE = L}
	\tilde{L}^\pm[(\Omega^\pm)^2 {g}_{\mu \nu}^\pm, \Omega^\pm, \psi^\pm, \partial_\alpha \psi^\pm]=(\Omega^\pm)^{-4}  {L}^\pm[{g}_{\mu \nu}^\pm, \psi^\pm, \partial_\alpha \psi^\pm],
\end{equation}
which is in agreement with \cite{Fleich} i.e.
\begin{equation} \label{g L TILDE = g L}
	\sqrt{g^\pm}\,  {L}^\pm[{g}_{\mu \nu}^\pm, \psi^\pm, \partial_\alpha \psi^\pm] \rightarrow \sqrt{\tilde{g}^\pm}\,\tilde{L}^\pm[{\tilde{g}}_{\mu \nu}^\pm, \psi^\pm, \partial_\alpha \psi^\pm].
\end{equation}
Similar statement is true for the other surface Lagrangians
\begin{equation} \label{Lagrangians}
	\begin{split}
			\tilde{\mathcal{L}}[\Omega^2 {h}_{\mu \nu}, \Omega, \psi, \partial_\alpha \psi]&=\Omega^{-3}  {\mathcal{L}}[{h}_{\mu \nu}, \psi, \partial_\alpha \psi] , \\
				\tilde{\mathfrak{L}}[\Omega^2 {h}_{\mu \nu}, \Omega]&=\Omega^{-3}  {\mathfrak{L}}[{h}_{\mu \nu}]. 
	\end{split}
\end{equation}
 Under (\ref{L TILDE = L}) and (\ref{Lagrangians}), the corresponding energy momentum tensors transform as
 \begin{equation} \label{EMTs}
 	\begin{split}
 		\tilde{\tau}_{\mu \nu}^\pm &=  (\Omega^\pm)^{-2}   	{\tau}_{\mu \nu}^\pm, \\
 			\tilde{S}_{\mu \nu} &=  \Omega^{-1}   	{S}_{\mu \nu},\\
 					\tilde{\Sigma}_{\mu \nu} &=  \Omega^{-1}   	{\Sigma}_{\mu \nu}.
 	\end{split}
 \end{equation}
   By setting $(\Omega^{\pm })^2= \Phi^\pm $ and $U^\pm(\Phi) = 2 \kappa (\Phi^{\pm })^2 V^\pm$, and considering (\ref{divergence}), equation (\ref{red dS tilde}) becomes
\begin{equation} 
	\begin{split}
			\delta	\tilde{S} &= \frac{1}{2 \kappa} \int_\mathcal{M^\pm} d^4x \sqrt{-g^\pm}\,   \bigg\{ 	\bigg( R^\pm - \frac{\partial  U^\pm}{\partial \Phi^\pm}  \bigg) \delta \Phi^\pm  + 2 \kappa  \bigg( \frac{\partial {{L}}^\pm}{\partial \psi^\pm} - \nabla_\alpha \frac{ \partial {{L}}^\pm}{\partial \partial_\alpha \psi^\pm} \bigg) \delta \psi^\pm    \\
		&+ \bigg(  \Phi^\pm G_{\mu \nu}^\pm+\frac{1}{2} U^\pm g^\pm_{\mu \nu} +\nabla_\alpha \nabla^\alpha \Phi^\pm g^\pm_{\mu \nu}-   \nabla_\mu \nabla_\nu \Phi^\pm -\kappa \tau_{\mu \nu }^\pm \bigg)      \delta g^{\pm \mu \nu} \bigg\} \\
		&-\frac{1}{2 \kappa}  \oint_\mathcal{\partial M} d^3y \epsilon \sqrt{|h|}  \bigg\{ \bigg(  \Phi \ [K_{\mu \nu}-K h_{\mu \nu} ] + n^\alpha \nabla_\alpha \Phi  \ h_{\mu \nu} +\kappa \epsilon   [{S}_{\mu \nu}+ {\Sigma}_{\mu \nu} ]\bigg) \delta h^{\mu \nu} - 2K  \delta \Phi \\
		&-2 \kappa   \bigg( \epsilon [ \frac{\partial {\mathcal{L}}}{\partial \psi} - {D}_\alpha \frac{ \partial {{\mathcal{L}}}}{\partial \partial_\alpha \psi}]+ {n}_\alpha \frac{\partial {{L}}}{ \partial \partial_\alpha \psi} \bigg)\delta \psi     \bigg\} -\frac{1}{2 \kappa} \int_\mathcal{\partial V^\pm} d^3y \epsilon \sqrt{|{h}|} \,   {{K}_0^\pm}  \bigg(\Phi^\pm {h}_{\mu \nu}^\pm  \delta {h}^{\pm \mu \nu}-3   \delta \Phi^\pm \bigg) \\
		&-\frac{1}{2 \kappa} \oint_{\partial \partial \mathcal{M}} d^2z \sqrt{{\sigma}}  \,   {N}_\alpha \bigg(\Phi {U}^\alpha + 2\kappa \epsilon \frac{ \partial {\mathcal{{L}}}}{\partial \partial_\alpha \psi}  \delta \psi \bigg).
	\end{split}
\end{equation}
 Although the above variation has the same bulk, corner, and surface terms regarding the matter field, it leads to different boundary equations of motion for induced metric\footnote{The variation of the non-dynamical term leads to different expression for $\delta \Phi$, however, leading to the same boundary equation.  }; However, the junction conditions for the latter would be the same. It is important to emphasize that the continuity of the scalar field indeed implies the continuity of the auxiliary field; Hence
  \begin{equation} \label{JC transformed}
 	\begin{split}
 		\Phi^+\big|_{\partial \mathcal{M}^+}&=\Phi^-\big|_{\partial \mathcal{M}^-}  \, ,\\ 
 		K^+ &=	K^- \, ,  \\
 	  \big[ K^+_{\mu \nu}-K^-_{\mu \nu} \big] \Phi \big|_{\partial \mathcal{M}}+ n^\alpha \big[ \nabla_\alpha \Phi^+ &-\nabla_\alpha \Phi^- \big]  h_{\mu \nu} = -\kappa \epsilon   \big[{\Sigma}^+_{\mu \nu}- {\Sigma}^-_{\mu \nu} \big]  \, .
 	\end{split}
 \end{equation}
 Here $S_{\mu \nu}=0$ has been used. Equation (\ref{JC transformed}) is exactly the same as previous junction conditions. In addition, one can can assume the continuity of the Ricci scalar as an extra condition in the Jordan frame; However this smoothness does not guaranteed in the Einstein frame, since by the second equation of (\ref{conformal Riemannian}),
 \begin{equation} \label{Ricci Einstein}
 	\begin{split} 
 	[\tilde{R}^+ - \tilde{R}^-]= -\frac{3}{\Phi^2|_{\partial \mathcal{M}}} [ \nabla_\alpha\nabla^\alpha \Phi^+ - \nabla_\alpha\nabla^\alpha \Phi^- ] , 
 	\end{split}
 \end{equation}
for $R^+ = R^-$. This non-vanishing term is a manifestation of different $V(\phi)$ or more precisely is the elegance of two different theories of gravitation.  
  
   \subsection{Invariance of action}
   Using the results of appendix and noting that the metric transfers with different conformal factor $\Omega^\pm$ in the sub-manifolds, we obtain
   \begin{equation} \label{transfered S_G}
   	\begin{split}
   		\tilde{S}_G[g]&=\frac{1}{2\kappa} \int_\mathcal{M^\pm} d^4x \sqrt{-g^\pm}\, \{ (\Omega^{\pm })^2 R^\pm - 6 \Omega^{\pm } \nabla_\alpha \nabla^\alpha \Omega^{\pm } \} -\frac{1}{ \kappa}  \oint_{\partial \mathcal{M}} d^3y \epsilon \sqrt{|h|} \ \Omega^2 (K - 3 n^\alpha \nabla_\alpha \ln \Omega ) +\frac{1}{ \kappa} \int_\mathcal{\partial V^\pm} d^3y \epsilon \sqrt{|{h}|} \, (\Omega^\pm)^2 {{K}_0^\pm}\\
   		&=\frac{1}{2\kappa} \int_\mathcal{M^\pm} d^4x \sqrt{-g^\pm}\, \{ (\Omega^{\pm })^2 R^\pm + 6 \nabla_\alpha \Omega^{\pm } \nabla^\alpha \Omega^{\pm } \} -\frac{1}{ \kappa}  \oint_{\partial \mathcal{M}} d^3y \epsilon \sqrt{|h|} \ \Omega^2 K+\frac{1}{ \kappa} \int_\mathcal{\partial V^\pm} d^3y \epsilon \sqrt{|{h}|} \, (\Omega^\pm)^2 {{K}_0^\pm},
   	\end{split}
   \end{equation}
   where we have used the integrating by parts in the second line.   By choosing $\phi^\pm= \sqrt{\frac{6}{\kappa}} \ln  \Omega^{\pm }$ the terms including the derivatives in the action cancels out each other, and considering the conformal transformation of the corresponding matter Lagrangians (\ref{L TILDE = L}) and (\ref{Lagrangians}), the total action reduce to
   \begin{equation} \label{invariance}
   	\begin{split}
   		\tilde{S}[g;\chi,\psi]&= \frac{1}{2 \kappa} \int_\mathcal{M^\pm} d^4x \sqrt{-g^\pm}\, \Big\{ \Phi^\pm R^\pm - U^\pm(\Phi^\pm) +2 \kappa {L}^\pm[{g}_{\mu \nu}^\pm, \psi^\pm, \partial_\alpha \psi^\pm] \Big\}\\
   		&-\frac{1}{ \kappa} \oint_{\partial \mathcal{M}} d^3y \epsilon   \sqrt{|h|} \Big\{ \Phi K +{\mathcal{L}}[{h}_{\mu \nu}, \psi, \partial_\alpha \psi] \Big\}+\frac{1}{ \kappa} \int_\mathcal{\partial V^\pm} d^3y \epsilon \sqrt{|{h}|} \, \Phi^\pm {{K}_0^\pm}.
   	\end{split}
   \end{equation}
   Here we have used $(\Omega^{\pm })^2= \Phi^\pm$ and   $U^\pm(\Phi^\pm)= 2\kappa (\Omega^{\pm })^4 V$. Therefore the action will be the same as the general $f(R)$ case. It is worthwhile to note that $\mathfrak{L}[{h}_{\mu \nu}]$ term is unnecessary and can be set to zero.

\section{Conclusion}
We have systematically derived the junction conditions for gluing different $f(R)$ gravitational theories using the variational approach. Our analysis reveals several key results:

In the Jordan frame, smooth junction conditions require:
\begin{equation} \label{JCs}
	\begin{split}
	\Phi^+\big|_{\partial \mathcal{M}^+}&=\Phi^-\big|_{\partial \mathcal{M}^-}  \, ,\\ 
K^+_{\mu \nu}&=K^-_{\mu \nu} \ , \\
 \nabla_\alpha \Phi^+ &=  \nabla_\alpha \Phi^- \ . 
	\end{split}
\end{equation}
In the Einstein frame, the equivalent conditions are:
\begin{equation} \label{JCs Einstein}
	\begin{split}
		\phi^+\big|_{\partial \mathcal{M}^+} &=\phi^-\big|_{\partial \mathcal{M}^-} \, ,\\
			\tilde{K}_{\mu \nu}^+  &=  \tilde{K}_{\mu \nu}^- \, ,  \\
		\tilde{\nabla}_\alpha \phi^+ &= \tilde{\nabla}_\alpha \phi^-  . \\
	\end{split}
\end{equation}
Several important insights emerge from our analysis. First, the continuity of the Ricci scalar $R$ across the boundary is not required for consistent matching, contrary to some expectations in modified gravity. Second, instead of the scalar curvature, the derivative $\partial f/\partial R$ must be continuous, imposing constraints on the functional forms of $f^+(R)$ and $f^-(R)$ at the junction.

Our work establishes that different $f(R)$ theories can be consistently joined provided their functional forms satisfy specific relations at the interface. The variational approach proves particularly powerful for this analysis, naturally handling the boundary terms essential for deriving consistent junction conditions.

Future work should address several important extensions in the theoretical framework: the null hypersurface case requiring modified boundary terms; dynamical boundary Lagrangians with independent gravitational degrees of freedom on the boundary; comparison with distributional approaches for higher-derivative theories and the junction conditions at infinity involving the $K_0$ counterterm. In addition to mathematical physics' interests, we think the studied framework is essential for some physical models. In this direction, we are  interested in some specific models like the special case of $R^2$ gravity relevant for re-normalized gravity \cite{R2} and \"uber-gravity  \cite{Khosravi:2017aqq,Khosravi:2016kfb} as well as symmetron models \cite{Hinterbichler:2010es}.  These investigations would further illuminate junction conditions in modified gravity and enable more sophisticated composite spacetime models.

\section*{Acknowledgments} \label{sec:ack}                                            
We would like to thank Reza Mansouri for his comments on this work.

\appendix
\section{Conformal transformation}
Under a conformal transformation
\begin{equation} \label{conformal g}
	\tilde{g}_{\mu\nu} = \Omega^2\, g_{\mu\nu},
\end{equation}
 which yields  $\sqrt{-\tilde{g}} = \Omega^{d} \sqrt{{-g}}$, where $d$ is the space-time dimension. For the unit normal vector $n_\mu$, the normalization condition $g^{\mu\nu}n_\mu n_\nu = \epsilon \quad$  with $ (\epsilon=\pm1)$ implies that 
 \begin{equation} \label{conformal normal}
 	\tilde{n}_\mu = \Omega\, n_\mu, \quad \text{and hence} \quad \tilde{n}^\mu = \Omega^{-1}\, n^\mu.
 \end{equation}
 The induced metric on a hypersurface transforms via
 \begin{equation} \label{conformal h}
 	\tilde{h}_{\mu \nu} = \Omega^2\, h_{\mu \nu}.
 \end{equation}
In addition, $\sqrt{\tilde{h}} = \Omega^{d-1} \sqrt{{h}}$, and $\tilde{h}_{\mu}^\nu=h_{\mu}^\nu$. The extrinsic curvature is defined by
\begin{equation} \label{Extrinsic curvature}
	K_{\alpha \beta} = -h_\alpha^{\ \mu}\, h_\beta^{\ \nu}\, \nabla_\mu n_\nu = - \nabla_\alpha n_\beta+\epsilon n_\alpha n^\mu \nabla_\mu n_\beta,
\end{equation}
transforms under \eqref{conformal g} as
\begin{equation} \label{comformal k}
	\tilde{K}_{\alpha \beta} = \Omega\left( K_{\alpha \beta} - h_{\alpha \beta}\, n^\mu \nabla_\mu \ln \Omega \right).
\end{equation}
Taking the trace with respect to the inverse induced metric, which transforms as $\tilde{h}^{\alpha \beta} = \Omega^{-2}\, h^{\alpha \beta}$, the transformed trace is
\begin{equation} \label{conformal trace}
\tilde{K} = \tilde{h}^{\alpha \beta}\, \tilde{K}_{\alpha \beta}
=\Omega^{-1}\bigg( K - (d-1)\, n^\mu \nabla_\mu \ln \Omega \bigg).
\end{equation}
Under conformal transformation, the standard Riemannian quantities transform as the following
\begin{equation} \label{conformal Riemannian}
		\begin{split}
					\tilde{R}_{\mu\nu} = R_{\mu\nu} - \bigg[ (d-2)\delta_\mu^\alpha \, \delta_\nu^\beta + g_{\mu\nu}g^{\alpha\beta} \bigg] \frac{\nabla_\alpha \nabla_\beta \Omega}{\Omega}  &+ \bigg[ 2(d-2)\delta_\mu^\alpha \, \delta_\nu^\beta - (d-3)g_{\mu\nu}g^{\alpha\beta} \bigg] \frac{(\nabla_\alpha \Omega)(\nabla_\beta \Omega)}{\Omega^2}, \\
				\tilde{R} = \frac{R}{\Omega^2} - 2(d-1)g^{\alpha\beta} \, \frac{\nabla_\alpha \nabla_\beta \Omega}{\Omega^3} &- (d-1)(d-4)g^{\alpha\beta} \frac{(\nabla_\alpha \Omega)(\nabla_\beta \Omega)}{\Omega^4}, \\
					\tilde{\nabla}_\alpha \tilde{\nabla}^\alpha S= \Omega^{-2} {\nabla}_\alpha {\nabla}^\alpha S &+ (d-2) \Omega^{-3}   g^{ \alpha \beta } {\nabla}_\alpha \Omega {\nabla}_\beta S.
		\end{split}
\end{equation}
For a vector field $A^\mu$, the conformal transformation of the divergence is
\begin{equation} \label{divergence}
	\tilde{\nabla}_\alpha \tilde{A}^\alpha = \frac{1}{\sqrt{-\tilde{g}}} \partial_\alpha (\sqrt{-\tilde{g}} \tilde{A}^\alpha ) = \frac{ \Omega^{-d}}{\sqrt{-{g}}} \partial_\alpha (\sqrt{-{g}}  \,\Omega^d \tilde{A}^\alpha )
\end{equation}
\section{Continuity of the fields across the boundary}
Any vector field $A_\mu$ can be decomposed as $ A_\mu = \epsilon n_\mu  n^\nu A_\nu + h_\mu^{ \, \, \nu} A_\nu$. In particular, the jump of the gradient of a scalar function $S$ can be decomposed as 
\begin{equation} \label{Jump of del S}
[	\nabla_\alpha S^+ - 	\nabla_\alpha S^-] = \epsilon n_\alpha n^\beta [\nabla_\beta S^+ - 	\nabla_\beta S^-] + h_\alpha^\beta \nabla_\beta [S^+ - S^-].
\end{equation}
The single-valued nature of $g_{\mu \nu}$ ensures that the variation of a one-parameter family of such a field, and each field in the family must be defined at every point on the manifold, including on the hypersurface
\begin{equation} \label{delta g}
	\begin{split}
		 \delta{g}_{\mu \nu}|_{\partial \mathcal{M}^+}=\delta{g}_{\mu \nu}|_{\partial \mathcal{M}^-},
	\end{split}
\end{equation}
this is due to the fact that $\delta g_{\mu \nu}|_{\partial \mathcal{M}^\pm}  = {g'}_{\mu \nu}|_{\partial \mathcal{M}^\pm} -g_{\mu \nu}|_{\partial \mathcal{M}^\pm}  $ and we have assumed that the metric is continuous across the boundary. The same statement is true for the induced metric. For the matter field $\psi$, the corresponding argument is not necessarily valid. In general, the matter field may have discontinuity across the boundary, i.e. $\psi|_{\partial \mathcal{M}^+} \neq \psi|_{\partial \mathcal{M}^-}$, implying 
\begin{equation} \label{delta psi}
\delta \psi|_{\partial \mathcal{M}^+} \neq  \delta \psi|_{\partial \mathcal{M}^-}. 	
\end{equation}
In a similar manner, the auxiliary field $\Phi^\pm$ or the scalar field $\phi^\pm$, may have discontinuity, and the above argument is valid in that case and these fields could have jump across the surface; but considering the scalar tensor theory,  relates the geometric quantities and physical fields (\ref{conformal g}) and  enforces continuous field since both conformal and ordinary metric must be continuous, hence   
\begin{equation} \label{Omega}
	{\Omega}|_{\partial \mathcal{M}^+}=  	{\Omega}|_{\partial \mathcal{M}^-},
\end{equation}
which yields
\begin{equation} \label{delta omega}
	\delta{\Omega}|_{\partial \mathcal{M}^+}=  	\delta {\Omega}|_{\partial \mathcal{M}^-}
\end{equation}
and by $\phi^\pm= \sqrt{\frac{6}{\kappa}} \ln  \Omega^{\pm }$, and $(\Omega^{\pm })^2= \Phi^\pm$ one can infer the same result for $\phi$ and $\Phi$ respectively. 

The surface energy momentum tensor in the \eqref{dS F}, can be written as
\begin{equation} \label{Sigma}
	\Sigma_{\mu \nu} \delta h^{\mu \nu} = - ( 	\Sigma^{\mu \nu}+2 \epsilon n^\mu n_\alpha 	\Sigma^{\alpha \mu} ) \delta h_{\mu \nu} - 2 (\epsilon n_\alpha 	\Sigma^{\alpha \nu}+ n^\mu n_\alpha n_\beta 	\Sigma^{\alpha \beta}  ) \delta n_\mu.
\end{equation}
The second term is the only term proportional to the variation of the normal vector, so by the least action principle, we demand it to be zero; $\epsilon n_\alpha 	\Sigma^{\alpha \mu}+ n^\mu n_\alpha n_\beta 	\Sigma^{\alpha \beta} =0$, taking the product of $n_\mu$ leads $n_\alpha n_\beta 	\Sigma^{\alpha \beta}=0$, so we conclude that 
\begin{equation} \label{Sigma n}
	n_\alpha	\Sigma^{\alpha \beta}=0,
\end{equation}
i.e. the boundary has no flux across its normal.

\section{Interpretation of equation (\ref{KG for PHI})}
Equation (\ref{KG for PHI}) can be regarded as a Klein-Gordon like of the auxiliary field i.e. a dynamical equation of motion for the third degree of freedom in $f(R) $ theory of gravity. This appendix is a proof of this statement. Starting from the second equation of (\ref{EOM bulk Einstein frame}) which is the Kelin-Gordon equation for the scalar degree of freedom, by using the last line of (\ref{conformal Riemannian}), the left hand side can be written as
\begin{equation} \label{Transf}
		\tilde{\nabla}_\alpha \tilde{\nabla}^\alpha \phi=
		\sqrt{\frac{3}{2\kappa}} \Phi^{-2} {\nabla}_\alpha {\nabla}^\alpha \Phi,
\end{equation}
 where we have used $\phi= 	\sqrt{\frac{3}{2\kappa}} \ln \Phi$. The first term in the right hand side of (\ref{EOM bulk Einstein frame})  can be reduced to
 \begin{equation}
 	-\frac{\partial \tilde{L}}{\partial \phi} =-\frac{\partial \tilde{L}}{\partial \Omega} \frac{\partial \Omega}{\partial \phi} = \tilde{\tau} \frac{\partial \ln \Omega}{\partial \phi} = 	\sqrt{\frac{\kappa}{6}} \Phi^{-2} \tau.
 \end{equation}
 Here, the chain rule has been used in the first equality and the imposing condition above the equation (\ref{imposed eq}) for the second. The last equality can be found by relating the traces of the first equation in (\ref{EMTs}). The second term in   (\ref{EOM bulk Einstein frame}) reduced to
\begin{equation}
	\frac{\partial V}{\partial \phi} = \frac{\partial \Phi}{\partial \phi} \frac{\partial V}{\partial \Phi}  =  \frac{\partial \Phi}{\partial \Omega} \frac{\partial \Omega}{\partial \phi} \frac{\partial V}{\partial \Phi} =   2 \Omega \frac{\partial \Omega}{\partial \phi}  \frac{\partial V}{\partial \Phi} =  2 \Omega^2 \frac{\partial  \ln \Omega}{\partial \phi} \frac{\partial V}{\partial \Phi}  = \sqrt{\frac{2 \kappa}{3}} \Phi \frac{\partial V}{\partial \Phi}.
\end{equation}
Since $V = \frac{1}{2 \kappa} \Phi^{-2} U= \frac{1}{2 \kappa} (\Phi^{-1} R -\Phi^{-2} f)  $, then $\frac{\partial V}{\partial \Phi} =\frac{1}{2 \kappa}  \Phi^{-3} (-R \Phi +2 f ) $, hence
\begin{equation}
		\frac{\partial V}{\partial \phi} = \sqrt{\frac{1}{6 \kappa}} \Phi^{-2} (-R \Phi +2 f ).
\end{equation}
Putting all together,  
\begin{equation}
		\tilde{\nabla}_\alpha \tilde{\nabla}^\alpha \phi  + \frac{\partial \tilde{L}}{\partial \phi}  - 	\frac{\partial V}{\partial \phi}= \sqrt{\frac{1}{6 \kappa}} \, \Phi^{-2}  \big( 3  {\nabla}_\alpha {\nabla}^\alpha \Phi   +R \Phi  - 2 f -\kappa\,\tau   \big),  
\end{equation}
Since the left hand side of the above equation vanishes, the desired equation can be verified,
\begin{equation}
	3  {\nabla}_\alpha {\nabla}^\alpha \Phi   + \Phi\,R  - 2 f =\kappa\,\tau.
\end{equation}

\end{document}